\documentclass[9pt]{article}
\usepackage{latexsym}
\usepackage{graphicx}
\usepackage{amsfonts,amssymb}

\begin{document}

\title{\textbf{Accuracy and Precision of Methods for Community Identification in Weighted Networks}}
\author{Ying Fan$^{1}$\footnote{Author for correspondence: yfan@bnu.edu.cn}, Menghui Li$^{1}$,
Peng Zhang$^{1}$, Jinshan Wu$^{2}$, Zengru Di$^{1}$\\
\\\emph{ 1. Department of Systems Science, School of Management,}\\
\emph{Beijing Normal University, Beijing 100875, P.R.China}
\\\emph{2. Department of Physics \& Astronomy, University of British Columbia, }\\ \emph{Vancouver, B.C. Canada, V6T 1Z1}}

\maketitle

\begin{abstract}
Based on brief review of approaches for community identification
and measurement for sensitivity characterization, the accuracy and
precision of several approaches for detecting communities in
weighted networks are investigated. In weighted networks, the
community structure should take both links and link weights into
account and the partition of networks should be evaluated by
weighted modularity $Q^w$. The results reveal that link weight has
important effects on communities especially in dense networks.
Potts model and Weighted Extremal Optimization (WEO) algorithm
work well on weighted networks. Then Potts model and WEO
algorithms are used to detect communities in Rhesus monkey
network. The results gives nice understanding for real community
structure.
\end{abstract}



\section{Introduction}

In recent years, more and more systems in many different fields
are depicted as complex networks, and recent empirical studies on
networks display that there are communities in social networks,
metabolic networks, economic networks
\cite{biological,InterbankMarket,metabolic,Nature,Bioinformatics}
and so on. As one of the important properties of networks,
community structure attracts us much attention.

In binary networks, community structure is defined as groups of
network vertices, within groups there are dense internal links
among nodes, but between groups nodes loosely connected to the
rest of the network\cite{Girvan}. Community structure is one of
the most important characters to understand the functional
properties of complex networks. For instance, in food web,
communities reveal the subsystem of ecosystem\cite{Williams}. In
the world wide web, the community analysis has found thematic
groups\cite{Eckmann,Lipowsky}. Email network can be divided into
departmental groups whose work is distinct and the communities
reveal organization structure\cite{emaila,emailb}. In biochemical
or neural networks, communities may correspond to functional
groups\cite{functional}. The deep understanding on community
structure will make us comprehend and analyze the characteristics
of systems better.

In order to partition communities of networks, many approaches
have been proposed recently
\cite{spectralbisection,KL,EPJB,HC,agent-based,spectral,
Ising,Wu,Donetti,Zhou,PRE,potts,EO,fast}. Most of these methods
detect communities according to topological structure or dynamical
behaviors of networks. Their sensitivity and computational cost
have been investigated by Danon et al recently\cite{compare}. In
practical situation, the number of communities are not known ahead
of time. So to evaluate the partition of networks quantitatively
is an important issue in community identification. Newman and
Girvan proposed a measurement for the partition of a network based
on topological structure, which is called modularity
$Q$\cite{PRE}. The value of modularity is lager, the partition is
better.

Most of above methods are developed for binary networks. In binary
networks, the closeness of relationships (link weight) among the
nodes are neglected. But in many real networks, link weight plays
an important rule in the structure and function of complex
networks. The community identification in weighted networks could
give us better understanding for the real functional groups.

For detecting community structure in weighted networks, the first
problem is how to define the community structure in weighted
networks. It is related with the evaluation of partitions of
weighted network. In binary networks, it is enough to consider
link density among nodes. But in weighted networks, the closeness
of relations also affect community structure. So the definition of
the community must integrate link with link weight. Newman has
generalized the modularity $Q$ to weighted modularity
$Q^w$\cite{weighted}. It suggests that we can depict the community
as follows: community structure is the groups of network vertices.
The fraction of internal link weight among nodes within groups is
much greater than that of link weight between groups. In other
words, the relations among nodes within groups are much closer
than relations between groups.

The second problem is that which method is more appropriate for
detecting community in weighted networks. Most approaches develope
for binary networks can be generalized to weighted networks. For
example, in presented algorithms, GN algorithm is based on the
link betweenees. When we take link weight into account, the link
betweenness can be gotten from weighted shortest path\cite{PRE}.
Potts model algorithm is based on Potts spin glass Hamiltonian,
and link weight could represent coupling strength between
spins\cite{potts}. Extremal Optimization(EO) method considers the
contribution of links to modularity\cite{EO}. We could generalize
it to Weighted Extremal Optimization(WEO) algorithm, which will
consider the contribution of link weight to weighted modularity.
Usually, groups separated with the link weight should be different
from the result based only on topological structure. The above
three methods give us the important examples of approaches for
community identification based on topological structure, dynamics,
and modularity respectively. In this paper, we mainly focus on GN
algorithm, Potts model and WEO method in weighted networks.

In order to find an appropriate approach on weighted networks, it
is necessary to evaluate accuracy and precision of each method. We
can apply each method to \emph{ad hoc} networks with a well known,
fixed community structure. Then accuracy can be got by comparing
the presumed communities and the communities found by the
algorithm. While precision can be calculated by comparing any pair
of communities found by the algorithm under same conditions.
Obviously, measuring the difference of different community
structures quantitatively is needed to evaluate accuracy and
precision. Newman described a method to calculate the sensitivity
of algorithms\cite{PRE}. Leon Danon et al proposed a measurement
$I(A,B)$ based on information theory\cite{compare}. These two
measurements mainly focus on the proportion of nodes which are
correctly grouped. In our previous work, we proposed a similarity
function $S$ to measure the difference between
partitions\cite{Dissimilarity}. It is based on the similarity and
dissimilarity of two sets and addresses the difference of number
of groups in different partitions. We will discuss these
measurements briefly and adopt similarity function $S$ to quantify
the difference of different partitions.

In this paper, we investigate the performance of several
generalized approaches for detecting community structures in
weighted networks. Both accuracy and precision are evaluated. The
presentation is organized as follows. In Section \ref{Approache},
we introduce the weighted modularity and some approaches for
community identification briefly. Then we review some methods to
measure the difference of different communities and introduce
similarity function $S$ to quantitatively describe the consistency
between different partitions. Section \ref{weightCommunity} gives
results of the accuracy and precision of several approaches based
on idealized binary and weighted networks. The results reveal that
weighted modularity $Q^w$ gives a nice description for the
community structure of weighted networks. Potts model based
approach and WEO approach work well in detecting community in
weighted networks. Then WEO approach is applied to Rhesus monkey
network. The result is consistent well with the real societies.
Some concluding remarks are given in Section \ref{conclud} .

\section{Weighted Modularity, Approaches and Measurements}\label{Approache}

\textbf{1. Weighted Modularity.} Link weight, as a strength of
interaction, is believed to be an important variable in networks.
It gives more information about the network besides its topology
dominated by links. Weighted networks can depict the real system
better and the community structure should take link weights into
account. So the definition of community\cite{Girvan} in weighted
networks need be modified.

Newman and Girvan have defined a modularity $Q$ to evaluate
community identification in binary networks\cite{PRE}. Newman have
indicated that modularity $Q$ can be generalized to weighted
networks\cite{weighted}:
\begin{equation}
Q^w=\frac{1}{2w}\sum_{ij}[w_{ij}-\frac{w_iw_j}{2w}]\delta(c_i,c_j),
\label{weightmodularity}
\end{equation}
where $w_{ij}$ is the link weight (similarity weight) between node
$i$ and $j$, $w_i=\sum_jw_{ij}$ (vertex weight) is the summation
of link weight attaching to node $i$,
$w=\frac{1}{2}\sum_{ij}w_{ij}$ is the summation of link weight in
the network, and $c_i$ shows that vertex $i$ belongs to community
$c_i$. Hence Eq. (\ref{weightmodularity}) can be write as
\begin{equation}
Q^w=\sum_{r}(e^w_{rr}-(a^w_{r})^2) \label{modularityweight}
\end{equation}
where
$e^w_{rr}=\frac{1}{2w}\sum_{ij}w_{ij}\delta(c_i,r)\delta(c_j,r)$
is the fraction of summation of link weight that connect two nodes
inside the community $r$,
$a^w_r=\frac{1}{2w}\sum_iw_i\delta(c_i,r)$ is the fraction of
summation of vertex weight of community $r$. Obviously, $Q^w$
takes both link and link weight into account. It suggests a
description for community in weighted networks. We think that
weighted community structure is the groups of network vertices.
The summation of internal link weight among nodes within groups
are larger than that of link weight between groups. In other
words, the relations of nodes within group are close, but the
relations of nodes between groups are distant. In the following
discussion, we evaluate a partition of weighted networks by the
weighted modularity $Q^w$ given by Eq. (\ref{modularityweight})
and $Q^w$ is considered as the global variable to optimize in
extremal optimization algorithm.

\textbf{2. Methods for Detecting Community.} In this paper, we
mainly discuss the the performance of GN algorithm, Potts model
algorithm, and Extremal Optimization algorithm in weighted
networks.

The GN algorithm is based on the concept of edge betweenness. The
betweenness of an edge in network is defined to be the number of
the shortest paths passing through it. It is very clearly that
edges which connect communities, as all shortest paths that
connect nodes in different communities have to run along it, have
a larger betweenness value. By removing the edge with the largest
betweenness at each step, the whole network can be gradually split
into isolated components or communities\cite{Girvan}. Based on
weighted shortest path, the above ideas can be easily generalized
to weighted networks\cite{weighted}.

For Potts model algorithm, community coincides with the domain of
equal spin value in the minima of modified Potts spin glass
Hamiltonian. The node in networks just be looked as the electron
with spin elected from $q$ spin states. The link correspond to the
reciprocity between electrons. This will allow us to partition the
communities of a network onto the magnetic domains in the ground
state or in local minima of a suitable Hamiltonian. For this
purpose authors\cite{potts} append a global constraint to the
q-state Potts Hamiltonian that forces the spins into communities
according to
\begin{equation}
H=\sum_{(i,j)\in
E}J_{ij}\delta_{\sigma_{i}\sigma_{j}}+\gamma\sum_{s=1}^{q}\frac{n_s(n_s-1)}{2}
\label{Hamilton}
\end{equation}
where, $\sigma_i$ denotes the individual spins which are allowed
to take $q$ values $1...q$, $n_s$ denotes the number of spins that
have spin $s$ such that $\sum^q_{s=1} n_s = N$, $J_{ij}$ is the
ferromagnetic interaction strength,
$\gamma=\frac{2<J_{ij}>M}{N(N-1)}$ is a positive parameter. To
practically find or approximate the ground state of system, a
simple Metropolis algorithm could be employed. For weighted
networks, $J_{ij}$ can be taken as similarity link weight. Then
Potts model algorithm can be easily applied to weighted networks.

The Extremal Optimization algorithm uses a heuristic search to
optimize the modularity $Q$ by dividing network into
groups\cite{EO}. When this method is generalized to weighted
networks, $Q$ is replaced by $Q^w$ as the global variable to
optimize. While the value of each node contributing to weighted
modularity $Q^w$ can be defined as
\begin{equation}
q^w_i=w_{r(i)}-w_ia^w_{r(i)}, \label{weightcontributing}
\end{equation}
where $w_{r(i)}$ is the summation of link weight that a node $i$
belonging to a community $r$ has with nodes in the same community,
and $w_i$ is the vertex weight of node $i$. So the modularity
$Q^w$ is $Q^w=\frac{1}{2w}\sum_iq^w_i$. Rescaling the local
variable $q^w_i$ by the vertex weight of node $i$, the
contribution of node $i$ to the weighted modularity is defined as
\begin{equation}
\lambda^w_i=\frac{q^w_i}{w_i}=\frac{w_{r(i)}}{w_i}-a^w_{r(i)},
\label{weightlamda}
\end{equation}
$\lambda^w_i$ is normalized in the interval $[-1,1]$. It gives the
relative contributions of individual nodes to the community
structure. So it could be considered as the fitness of a node
involved in the weighted extremal optimization process. The
process of detecting community structure by weighted extremal
optimization (WEO) is as follows:
\begin{enumerate}
\item Initially, split randomly the whole network into two groups
with similar number of nodes. \item At each time step, move the
node with the lower fitness from one group to the other. After
each movement, recalculate the fitness of every node based on Eq.
(\ref{weightlamda}). \item Repeat process $2$ until a maximum
value of $Q^w$ is reached. After that, proceed recursively with
every group. When the modularity $Q^w$ cannot be improved, the
process will finish.
\end{enumerate}
In order to escape from local maxima, WEO algorithm adopts
$\tau$-EO method\cite{tau}. The node is selected according to the
following probability:
\begin{equation}
P(q)\propto q^{-\tau},
\end{equation}
where $q$ is the rank number of node according to their fitness
values, and $\tau \sim 1+1/ln(N)$.

\textbf{3. Comparing Two Community Structures.} In order to find
appropriate method to solve a certain problem, we should know the
performance of each method, including their speed and sensitivity.
Usually the sensitivity of a algorithm is tested by its
performance when it is applied to \emph{ad hoc} networks with a
well known, fixed structure. The previous researches focus only on
the accuracy of a method. Here we address that both accuracy and
precision of a algorithm should be examined. When we proceed the
methods several times under the same condition, they may give
different community structures due to the random factors in the
algorithm. Instead of comparing with the presumed communities to
get the accuracy, the precision should be got by comparing the
results from different runs under the same conditions. Obviously,
in order to get accuracy or precision of a method, we need to
quantify the differences between different community structures.
In other words, we need a measurement to evaluate the similarity
between communities.

There are already several methods to quantify the difference of
different partitions. Newman described a evaluating method in
\cite{PRE}. The largest set of vertices grouped together in each
of the four known communities is considered correctly classified.
If two or more of known sets are put into the same group, then all
vertices in those sets are considered incorrectly classified. All
other vertices not in the largest sets are considered incorrectly
classified. Leon Danon et al proposed the use of the
\emph{normalized mutual information} measure\cite{compare}. It is
based on the \emph{confusion matrix} $N$, where the rows denote
the presumed communities before divided, and the columns
correspond to the communities found by some algorithm. The matrix
element($N_{ij}$) of $N$ is the number of nodes in the presumed
community $i$ that appear in the found community $j$. A measure of
similarity between the partitions, based on information theory, is
then:
\begin{equation}
I(A,B)=\frac{-2\sum^{c_A}_{i=1}\sum^{c_B}_{j=1}N_{ij}\log(\frac{N_{ij}N}{N_{i.}N_{.j}})}
{\sum^{c_A}_{i=1}N_{i.}\log(\frac{N_{i.}}{N})+\sum^{c_B}_{j=1}N_{.j}\log(\frac{N_{.j}}{N})}
\label{mutual}
\end{equation}
where $c_A$ is the number of presumed communities and $c_B$ is the
number of found communities, $N_{i.}$ is the sum over row $i$ of
matrix $N_{ij}$ and $N_{.j}$ is the sum over column $j$.

The above two measurements mainly focus on the proportion of nodes
which have been grouped correctly. We have also suggested a method
to characterize the difference of community structures
quantitatively\cite{Dissimilarity}. It is based on the similarity
and dissimilarity of two sets $A$ and $B$ defined as the subset of
$\Omega=A\bigcup B$. The normalized similarity are defined as
\begin{equation}
s = \frac{\left|A\cap B\right|}{\left|A\cup B\right|}\\
\end{equation}
Given any two partitions, we should first construct correspondence
between two subsets of two partitions. Then we can get similarity
function $S$ by integrating the results of every single
pair\cite{Dissimilarity}.

The former two methods evaluate the sensitivity of algorithm by
measuring the percentage of nodes divided correctly, while the
similarity function $S$ we proposed emphasize the difference of
communities. In our method, all clusters have equal status
regardless of its size. For a network with 128 nodes and 4
presumed groups with 32 nodes each, if there are three communities
divided by a method, two of which correspond exactly to two
original communities, and a third, which corresponds to the other
two clustered together, the similarity is $S=0.625$, Newman's
measurement gives 0.5, and $I(A,B)=0.858$. Considering another
example, a network consists of $n=20$ vertices and is divided into
two presumed groups of $10$ nodes each. It is provided by a method
that there are three communities in the result. The two largest
groups are divided correctly except one node which forms the third
group. In this case, the accuracy given by each measurements are
$0.95$(Newman), $0.9$ ($I(A,B)$), and the similarity function is
$0.63$. So our method for quantifying the difference of partitions
emphasizes the number of communities. It is a reasonable
evaluating index to quantify the difference of community
structures. In the following discussion, we use $S$ to describe
the accuracy and precision of the approaches.

\section{Results Based on Idealized Networks and Empirical Studies}\label{weightCommunity}

The above methods will be applied to \emph{ad hoc} networks
firstly introduced by Newman and used by many other
authors\cite{PRE,compare}. Each network consists of $n=128$
vertices, which divided into four groups of $32$ nodes. Vertices
are assigned to groups and are randomly connected to vertices of
the same group by an average of $\langle k_{in}\rangle$ links and
to vertices of different groups by an average of $\langle
k_{out}\rangle$ links. The average degree of all vertices are
fixed, namely $\langle k_{in}\rangle+\langle k_{out}\rangle=16$.
With $\langle k_{out}\rangle$ increasing from small, the
communities become more and more diffuse, and it becomes more and
more difficult to detect the communities. Since the community
structure is well known in this case, it is possible to measure
accuracy and precision of each method by quantifying the
difference of partitions.

For each approach of community identification, accuracy could be
gotten by the comparison between the divided communities with the
presumed one and precision should be gotten by the comparison
between any pair of results, which are found by same algorithm
performing several times on the same network. In the following
numerical investigations, we first get $20$ realizations of
idealized \emph{ad hoc} networks under the same conditions. Then
we run each algorithm to find communities in each network $10$
times. Based on these results, using the similarity function $S$,
comparing each pair of these $10$ community structures and average
over the $20$ networks (average of $900$ results) could give us
the precision of the algorithm. Comparing each divided groups with
the presumed structure, we can get the accuracy of the algorithm
by averaging these $200$ results.

\subsection{Results for binary networks}

\begin{figure}
\includegraphics[width=6cm]{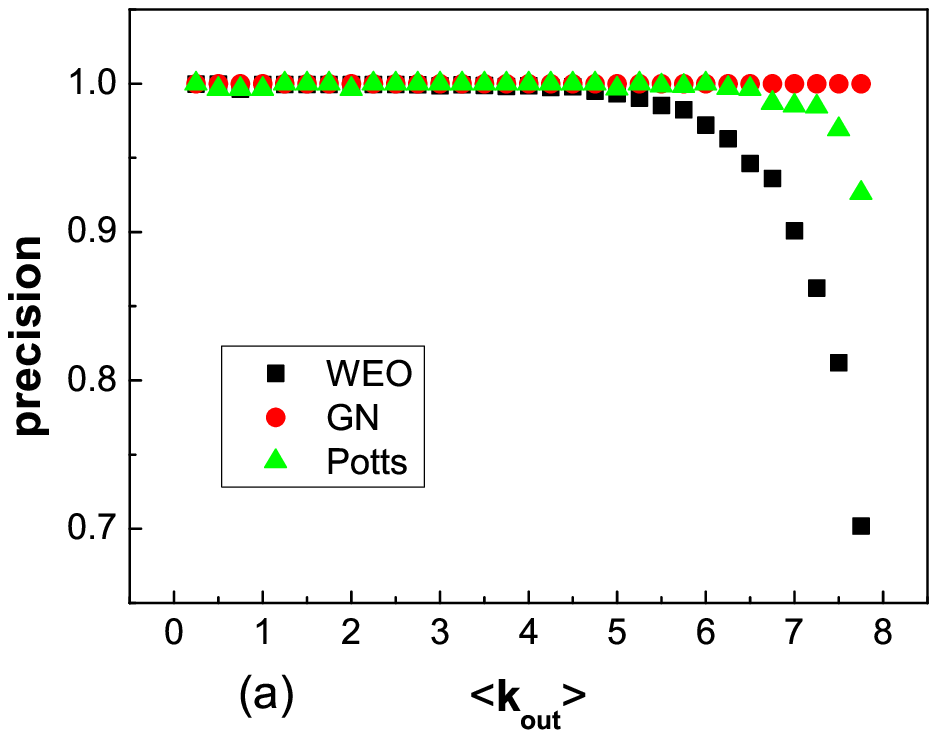} \includegraphics[width=6cm]{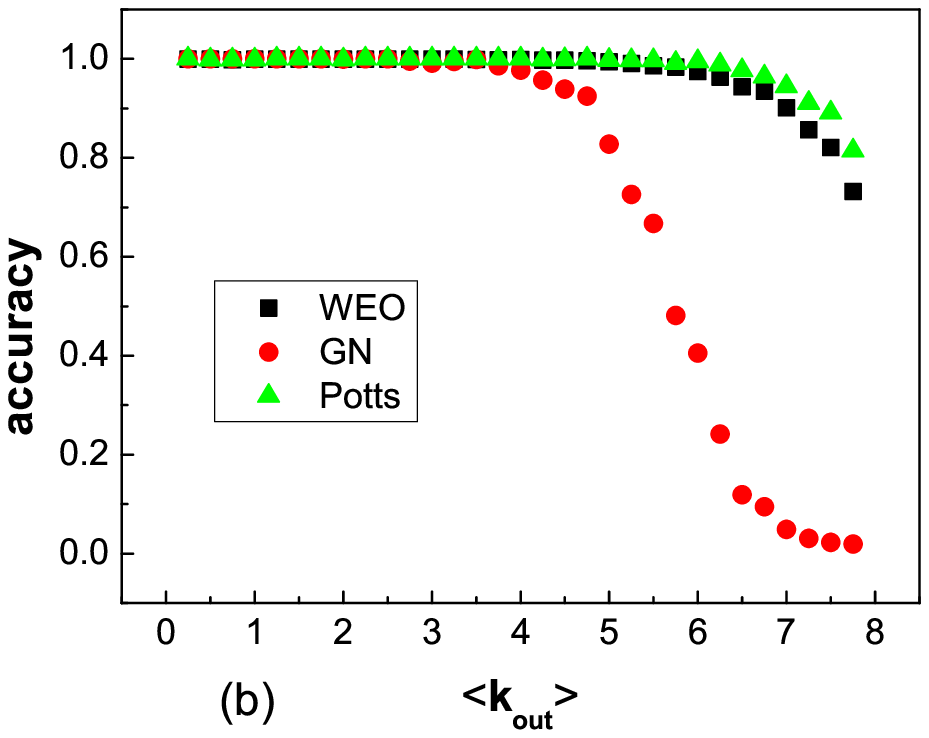}
 \caption{Algorithm performance as applied to \emph{ad hoc} networks with $n=128$ and four
communities of 32 nodes each. Total average degree is fixed to 16.
(a)Comparing precision of the algorithm by several results on same
\emph{ad hoc} networks. (b)Comparing accuracy using \emph{ad hoc}
networks with presumed community structure. The \emph{x}-axis is
the average of connections to outside communities $\langle
k_{out}\rangle$. Each point is an average over $20$ networks and
$10$ runs each.}\label{accuary}
\end{figure}

We apply GN, Potts model, and WEO algorithm on binary \emph{ad
hoc} network first and focus on both accuracy and precision of the
methods measured by similarity function $S$. From the results
shown in Fig.\ref{accuary}, when $\langle k_{out}\rangle$ is
small, they could find communities well and truly. The communities
become more diffuse with $\langle k_{out}\rangle$ increasing. Once
$\langle k_{out}\rangle$ is lager than a certain value, it is
difficult to find presumed communities exactly. For small $\langle
k_{out}\rangle$, there are no discrimination between results of
any algorithms. But for large $\langle k_{out}\rangle$, the
precision of different algorithms is various
(Fig.\ref{accuary}(a)) and accuracy falls across the different
algorithms. For example, GN algorithm is stable for any $\langle
k_{out}\rangle$, though its accuracy is worse when $\langle
k_{out}\rangle$ is large. However, Potts model and EO algorithm
are fluctuant, though their accuracy are better.

As shown in Fig.\ref{accuary}(b), with $\langle k_{out}\rangle$
increasing, accuracy of GN algorithm described by similarity
function $S$ decline more quickly than the other measurements. It
is because there are many small clusters in found communities by
GN algorithm and similarity function $S$ addresses the number of
communities crucially.

\subsection{Results based on weighted \emph{ad hoc} networks}

In this section, \emph{ad hoc} networks is added similarity link
weight to describe the closeness of relations. Similarity weight
is proportional to the closeness of relationships. The larger the
link weight is, the closer the relation is. Under the basic
construction of \emph{ad hoc} network described above, the weight
of link connected to vertices of the same group is assigned as
$1$, while the weight of link connected to vertices of different
groups is assigned as $w_{out}$. In practise, the relationship
among the nodes in groups is usually more closer than the
relationship between groups. So $w_{out}$ is normally less than 1.
When $w_{out}$ is equal to $1$, weighted \emph{ad hoc} network
could be seen as a binary one.

\begin{figure}
\includegraphics[width=6cm]{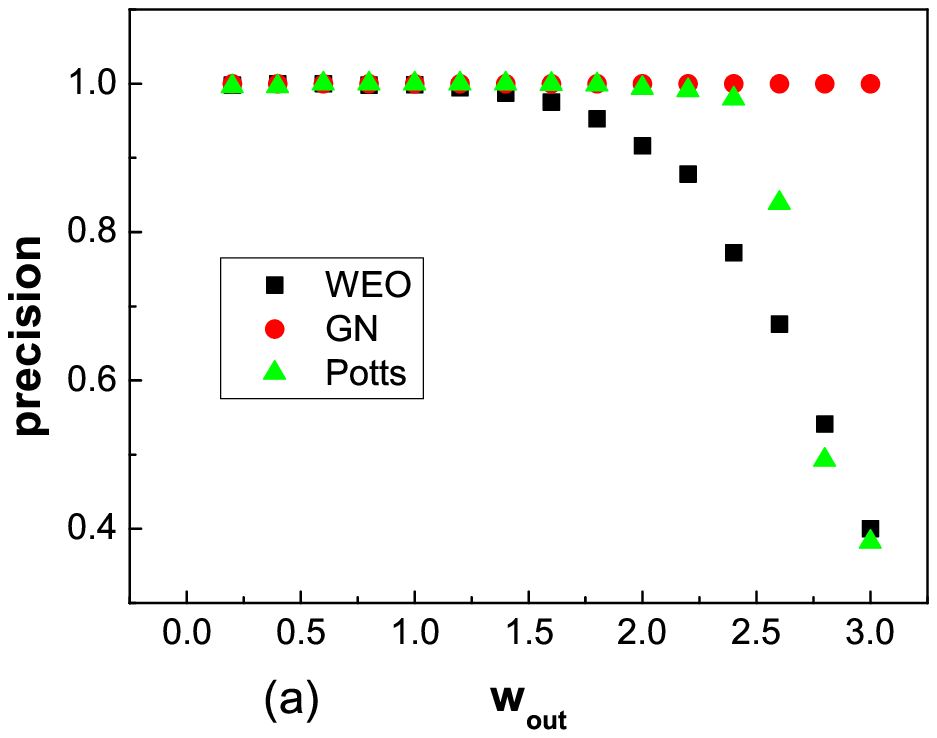} \includegraphics[width=6cm]{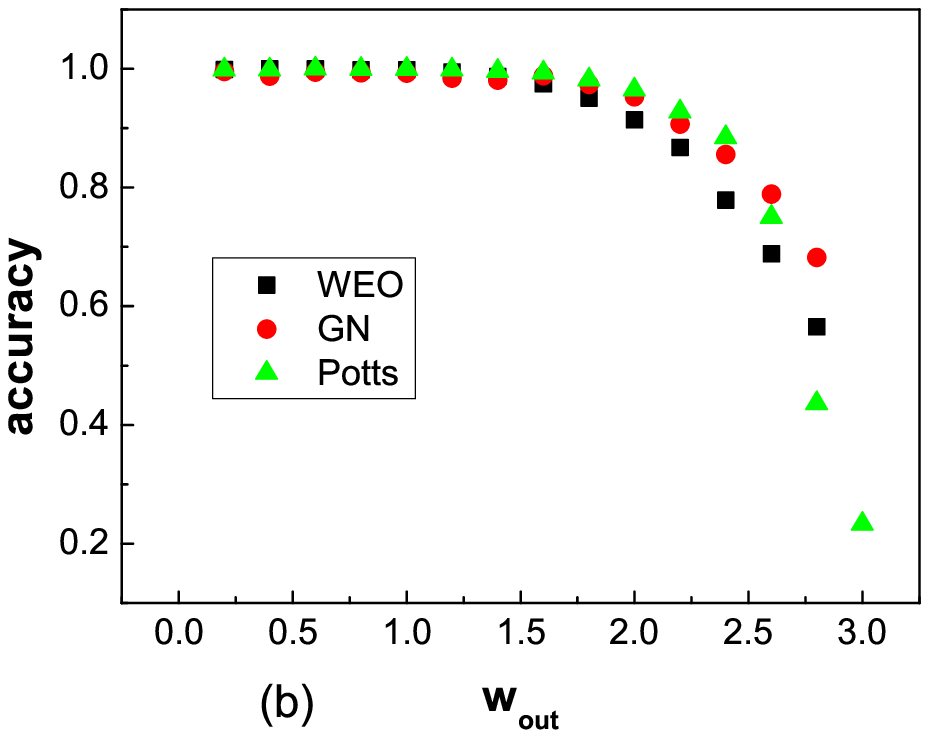}
\caption{Networks with $\langle k_{out}\rangle$ equaling to $4$.
The link weights among the nodes in groups are fixed to 1. The
influence of link weight on the out edge on precision(a) and
accuracy(b) when out link weight changes from $0.2$ to $3$. Each
point is an average over $20$ networks and $10$ runs
each.}\label{weight12-4}
\end{figure}

We consider weighted modularity $Q^w$ as the criteria to
evaluating partition of communities. Though there are only two
kinds of link weight value in weighted \emph{ad hoc} networks, the
role of weight on community structure can be investigated
qualitatively. When $\langle k_{out}\rangle$ is fixed, with
$w_{out}$ increasing from small, it affects community structure
obviously.

When $\langle k_{out}\rangle$ is equal to $4$, any methods can
easily partition communities for binary networks. When $w_{out}$
is small or even equals 1, all algorithms works well to get the
correct communities (as shown in Fig.\ref{weight12-4}). When
$\langle k_{out}\rangle$ is small, we know that the community
structure is dominated by links. So we can find the presumed
groups even when $w_{out}$ is larger than 1.

\begin{figure}
\includegraphics[width=6cm]{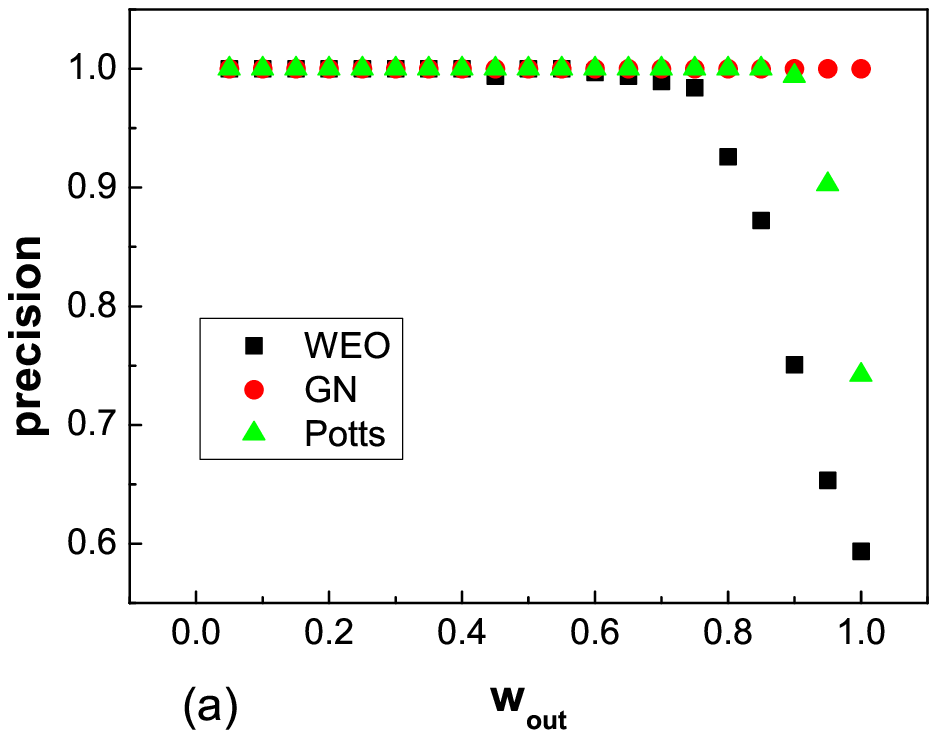}\includegraphics[width=6cm]{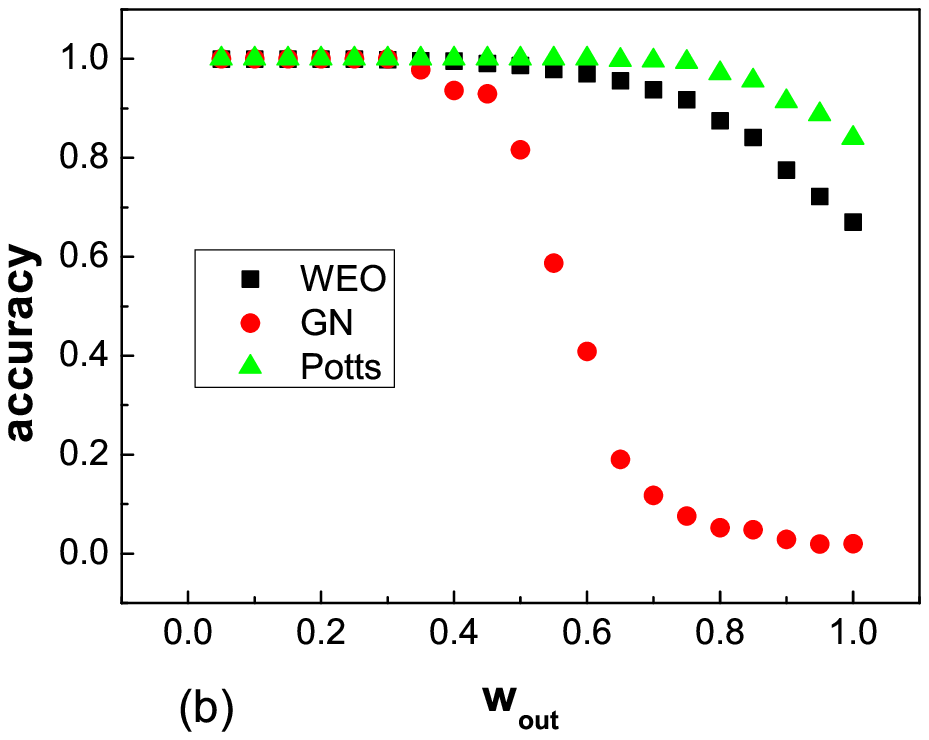} \caption{The
influence of weight on community structure when $\langle
k_{out}\rangle$ is equal to $8$. A series of networks possess
identical topological structure, and out link weight changes from
$0.05$ to $1$. Each point is an average over $20$ networks and
$10$ runs each.}\label{weight-8-8}
\end{figure}

In other hand, when $\langle k_{out}\rangle$ is large, the
communities is very diffuse in binary networks and it is
impossible to find communities correctly by any algorithms. But
now the link weight plays a crucial role in the partition of
communities. When $w_{out}$ is small, the network can also be
partitioned accurately into presumed communities (shown as
Fig.\ref{weight-8-8}). In this case, Potts model and WEO algorithm
work better than GN algorithm. Although GN algorithm has nice
precision, it gives results in low accuracy.

From the above results, we can see that topology and link weight
are two factors that affect community structure. When we set
$w_{out}$ equal 0.2, Fig.\ref{weight-0.2} shows the precision and
accuracy of Potts model and WEO algorithm as the function of
$\langle k_{out}\rangle$. The spectrum of $\langle k_{out}\rangle$
is lager than that of binary networks. The results show the link
weight really plays an important rule in community structure and
the Potts model and WEO methods are effective in detecting
communities in weighted networks. Even in complete networks, when
we set the link weight be $1$ among the same presumed group, while
the link weight is $w_{out}$($w_{out}<1$) between groups, these
two algorithms can divided the network into communities correctly
(as shown in Fig.\ref{complet}). In dense networks, link weight is
more important than in sparse networks. So weighted modularity
$Q^w$ and corresponding algorithms are helpful to detecting
community structure in dense weighted networks.

\begin{figure}
\includegraphics[width=6cm]{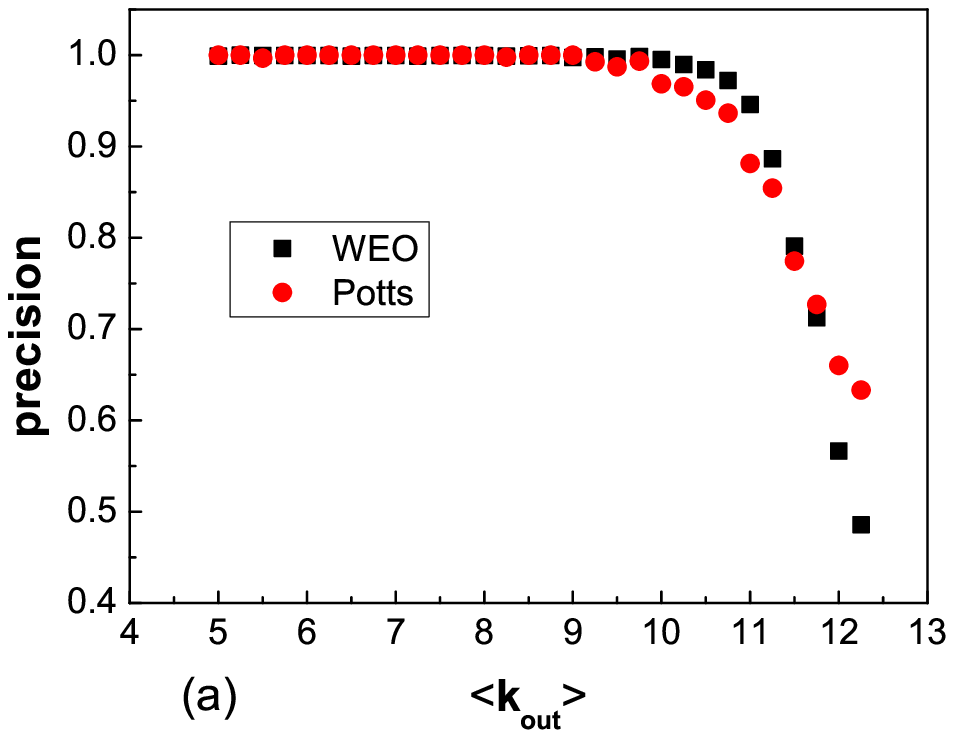}\includegraphics[width=6cm]{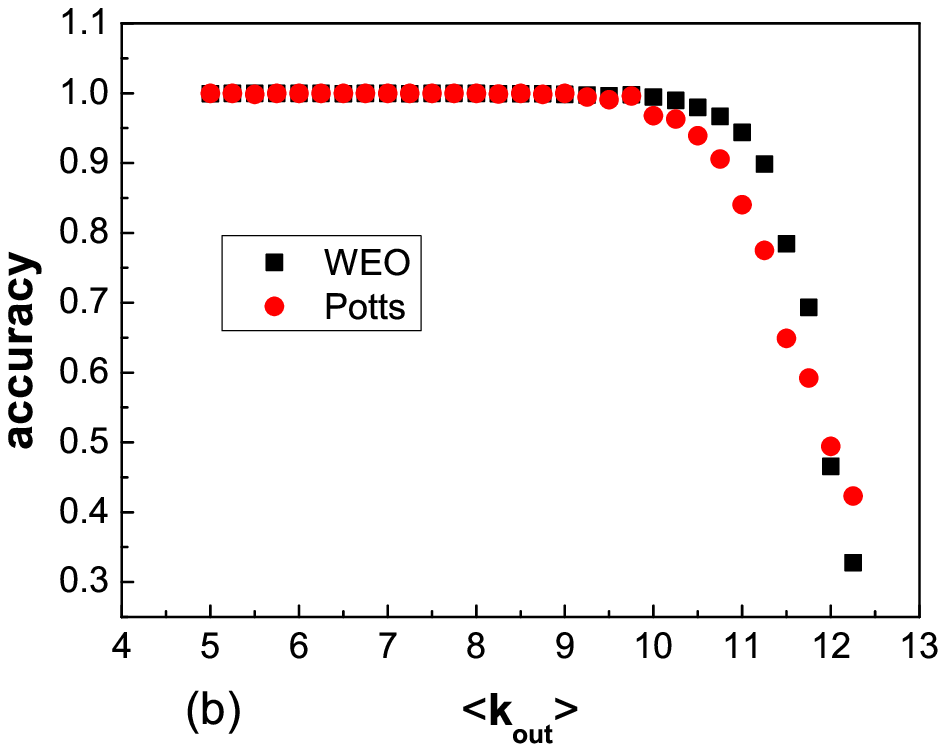}
\caption{The influence of topology on community structure when
$w_{out}$ is fixed as $0.2$. Each point is an average over $20$
networks and $10$ runs each.}\label{weight-0.2}
\end{figure}

\begin{figure}
\center
\includegraphics[width=6cm]{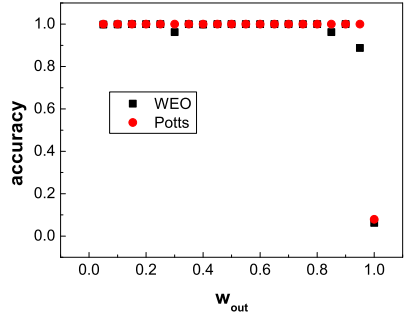}
\caption{Accuracy of Potts and WEO algorithms in complete networks
with presumed communities. When $w_{out}$ is less than 1, the
algorithms can find the groups correctly. Each point is an average
over $10$ runs.}\label{complet}
\end{figure}

\begin{figure}
\includegraphics[width=6cm]{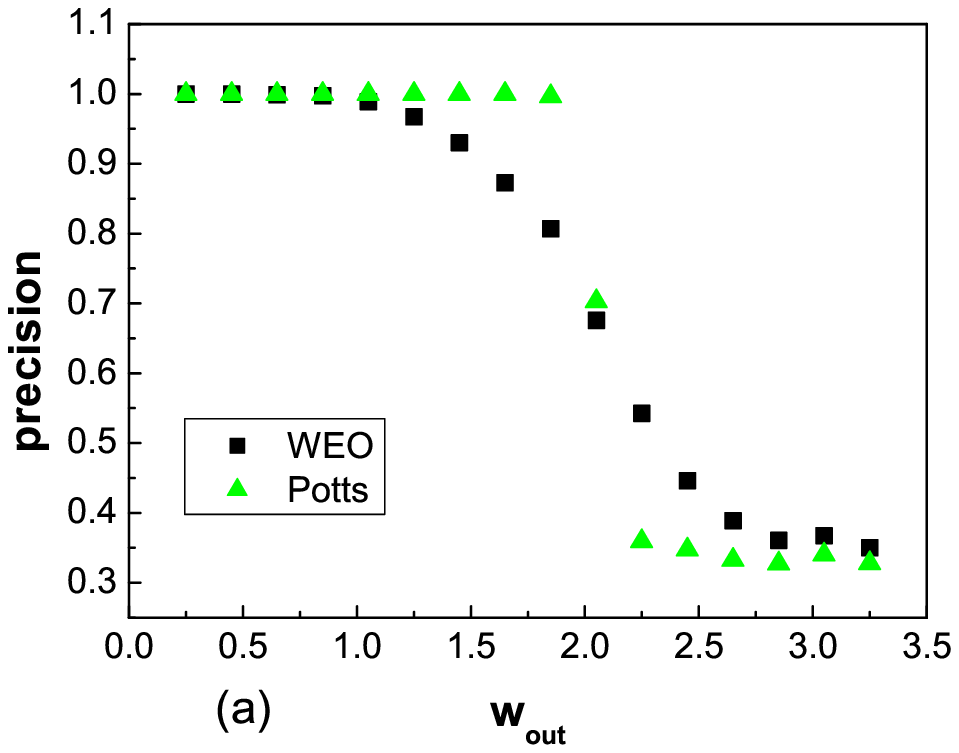}\includegraphics[width=6cm]{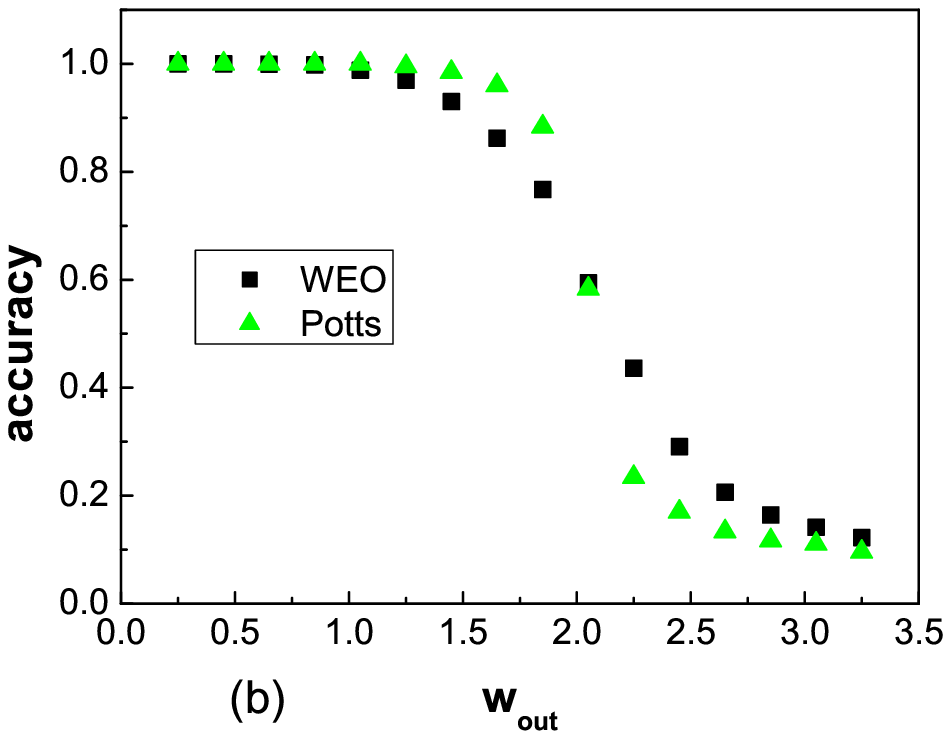}
\includegraphics[width=6cm]{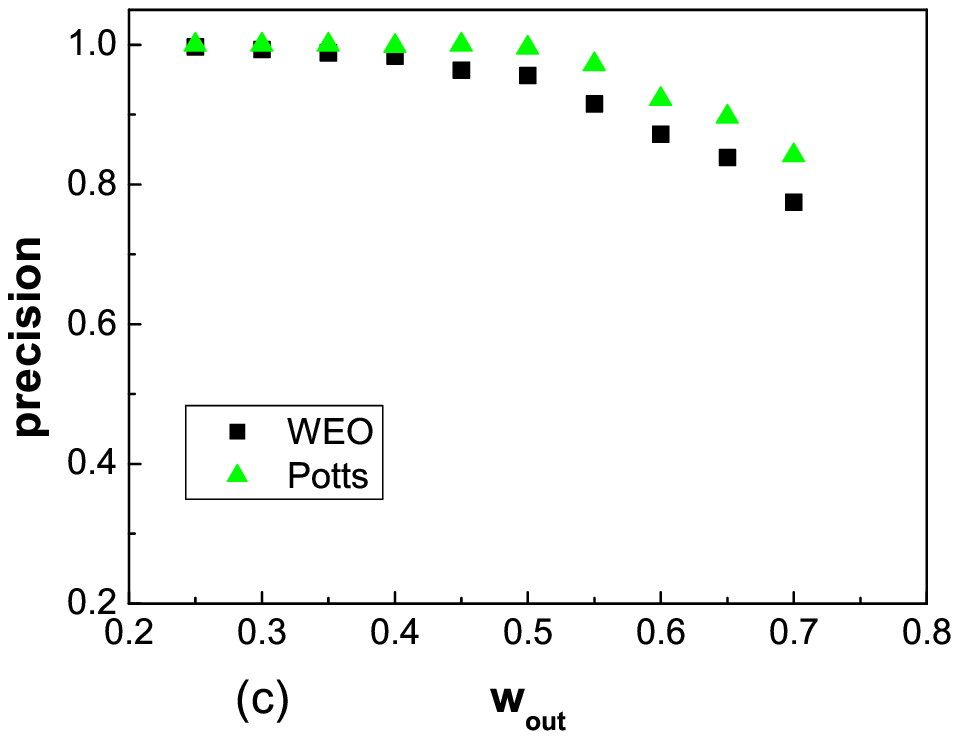}\includegraphics[width=6cm]{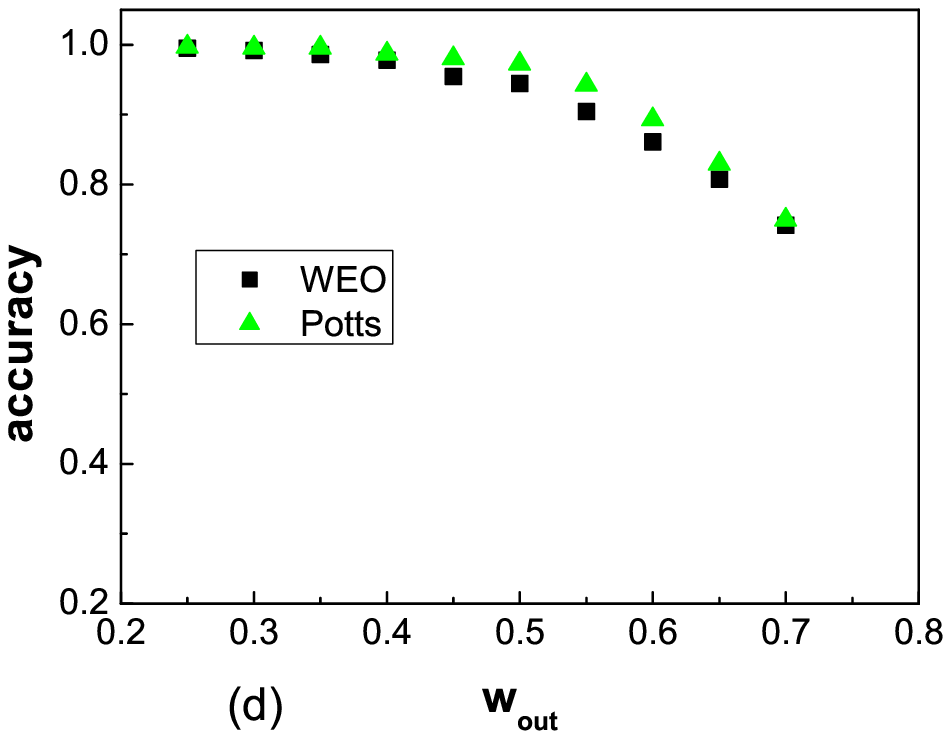}
\caption{The performance of Potts model and WEO algorithms in
networks with randomly distributed link weights. (a) and (b) are
the precision and accuracy of the algorithms when $\langle
k_{out}\rangle$ is equal to $4$. (c) and (d) are the results when
$\langle k_{out}\rangle$ is equal to $8$. Each point is an average
over $20$ networks and $10$ runs
each.}\label{weight-0.2}\label{random}
\end{figure}
In real weighted networks, link weights are usually randomly
distributed. We have also tested Potts model and WEO algorithm in
the idealized \emph{ad hoc} networks with random link weight
distribution. For a given network topology with certain $\langle
k_{out}\rangle$, weights in links among groups are taken randomly
from [0.5,1]. The average link weight in groups is 0.75. While
weights in links between groups ($w_{out}$) are also taken
randomly from a interval with the same length. With the changing
of its average value, we can also get the performance of
algorithms under different conditions. The results are summarized
in Fig.\ref{random}. They are qualitatively similar with the above
results.

\subsection{Community structure in Rhesus monkey network}

The above investigations are based on idealized networks. Now we
move to some real weighted networks. One example is Rhesus monkey
network which is studied by Sade\cite{rhesus} in $1972$. This
network is based on observations of group F Cayo Santiago, in
which $38$ monkeys comprise $6$ genealogies and $2$ non-natal
males ($066$ and $R006$). The grooming episodes of monkeys were
registered between $14th$ June and $31st$ July $1963$, just prior
to the mating season on Cayo Santiago. The network showed the
information of members, who were $4$ years old or older in $1963$.
Links denote grooming behavior between the monkeys and link weight
is the number of instances of grooming of each monkey by each
other during the period of observation. The network has $16$
vertices and $69$ links with link weights ranging from $1$ to
$49$.

\begin{figure}
\includegraphics[width=6cm]{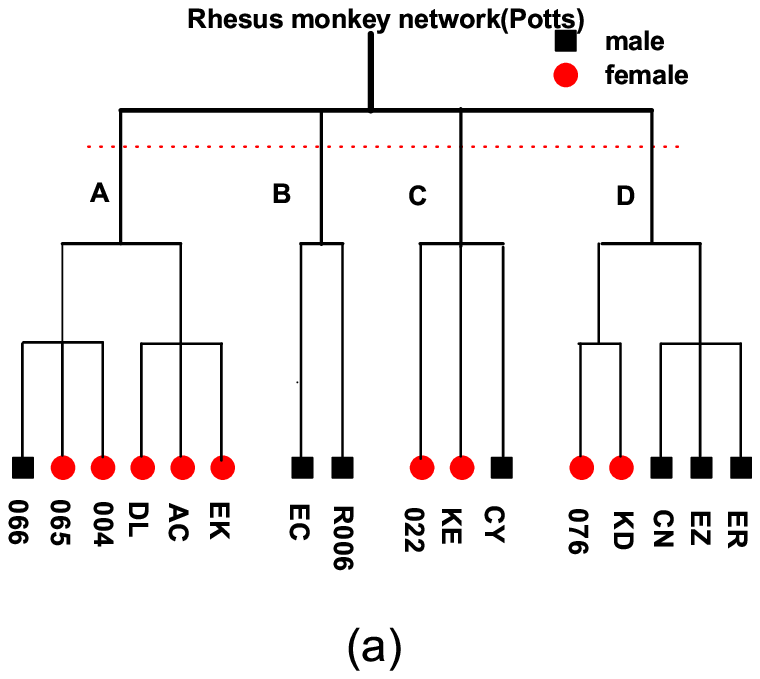}\includegraphics[width=6cm]{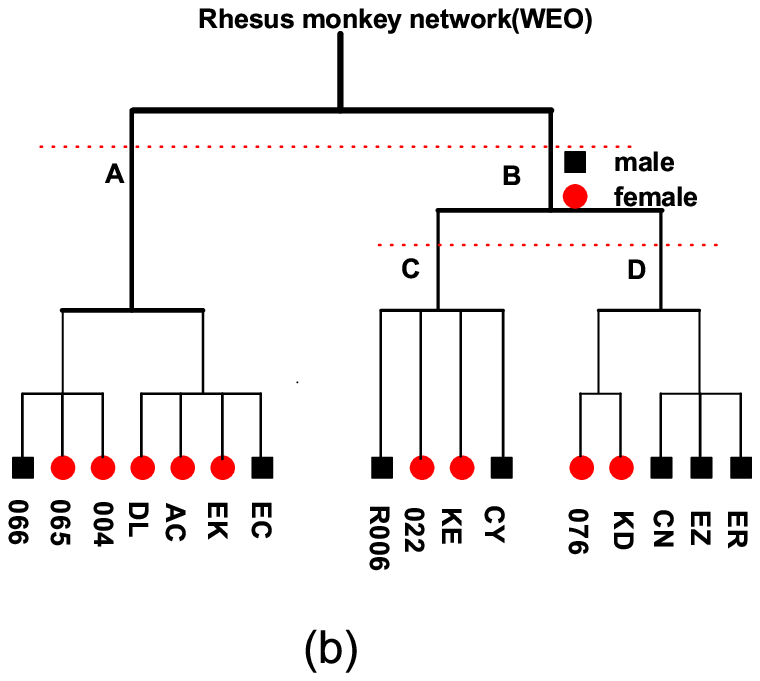}\caption{(a)Community
structure of rhesus monkeys network gotten by Potts model. It is
the result occurs with maximum probability. (b) Groups gotten by
WEO algorithm. $DL$ and $EC$ were $AC$'s offspring, $CY$ and $KE$
were $022$'s offspring, and $EZ$ and $ER$ were
brothers.}\label{monkey}
\end{figure}

Newman has illuminated its community structure by weighted GN
algorithm (See Fig.2 (b) in \cite{weighted}). Here we test Potts
model and WEO algorithm in Rhesus monkey network. Two algorithms
have been applied 20 times to get the communities. Their precision
measured by similarity function are: 0.95(Potts) and 1.00(WEO)
respectively. The final community structures are shown in
Fig.\ref{monkey}. Fig.\ref{monkey}(a) shows the result gotten by
Potts model algorithm. The network is partitioned into 4 groups.
Its $Q^w$ is 0.23. By WEO algorithm, the network is divided into
three communities(A, C and D). Firstly, rhesus monkey network is
partitioned into two groups(A and B), then group B is divided into
two smaller groups (Fig.\ref{monkey}), and the max weighted
modularity $Q^w$ is $0.244$. $Q^w$ for partition (4 groups) in
\cite{weighted} gotten by GN method is 0.12. We have applied WEO
algorithm further and have divided group $A$ and $D$ into two
smaller groups. The community structures gotten by Potts and WEO
algorithm are different from the results gotten by GN methods. The
similarity of these communities given by similarity function $S$
are: 0.65(WEO vs. Potts), 0.22(WEO vs. GN), and 0.36(Potts vs.
GN). By the detailed investigation, it could be found that results
gotten by WEO and Potts model methods accord more details of the
known organization of these monkeys.

From the record by Sade\cite{rhesus}, Male $006$ had been dominant
in group since at least $1960$. Male $R006$ had been solitary in
$1962$, and joined this group in early $1963$. $R006$ replaced
$006$ as dominant male in the fall of $1963$.
Based on the investigation of cliques, the following details can
be known. The dominant male $066$ was co-cliqual with the first
and second dominant females. $EC$, a 4-year-old male, is
co-cliqual with his mother $AC$ and sister $DL$. The $4$
multi-cliqual monkeys, $065$, $004$, $AC$ and $DL$, were the $4$
highest ranking females and they formed the core of the grooming
network. (This can be reflected by community A.) $EZ$ occurs only
in his brother $ER$'s clique. $CN$, the adult male castrate, is
co-cliqual with his mother and sister, overlapped more extensively
with the cliques containing the other natal males (community D can
show this detail), and might link their clique to the main group.
$R006$, the new-natal male, at last did not clique with the
dominant and third ranking females, $065$ and $AC$ and $AC$'s son
$EC$. (This can be reflected by community C.) In conclusion, the
dominant male $066$, was integrated into the core with the
females. The new male, $R006$, was distantly attached to the core
of females. one sub-adult male, $EC$, was still integrated into
his genealogy. The other natal males formed a distinct sub-group.
$CN$, the castrate, was intermediate in his position, which
overlapped that of the natal males and the female core. The
community structure found by WEO algorithm can illuminate above
details well. This example suggests that WEO algorithm is
effective at finding community structure in weighted networks.

\section{Concluding Remarks}\label{conclud}

In this paper, we focus on the identification of community
structure in weighted networks. When link weight is taken into
account, the closeness of relations among the nodes in a group
should be characterized both by link and link weight. Weighted
modularity $Q^w$ should be taken as the criteria for partition of
communities. A brief overview of some partition algorithm is given
including the introduction of local variable $\lambda ^{w}_i$ of
weighted network in the external optimization process. In
addition, we review the methods to evaluate accuracy and precision
of different algorithms and present similarity function $S$, a
measurement to quantify the difference of different communities.
In weighted \emph{ad hoc} networks, we study the influence of
weight on the process of detecting community structure and find
that the change of link weight can affect the accuracy and
precision of algorithms. When community structure is dominated by
topological linkage, GN algorithm works better than other
algorithms. But for dense networks, when link weight plays more
important rule in network properties, Potts model and WEO
algorithm works better to get the correct communities. At last, we
use WEO algorithm to Rhesus monkey network. The results show that
community identification with weighted modularity $Q^w$ and WEO
algorithm gives better understanding for real networks.

From these investigations, we could find that the role of weight
on the weighted networks could be investigated by studying the
effect of weight on the community structure. For weighted
networks, the disturbing of distribution or matching between
weights and edges should have some important effects on community
structure. So the community structure in networks should be a
suitable property for investigating the role of weight.

\section*{Acknowledgement}
The authors want to thank Dr. Newman for his cooperation data.
This work is partially supported by 985 Projet and NSFC under the
grant No.70431002, No.70371072 and No.70471080.

\end{document}